\documentclass[english,aps,pra,reprint,superscriptaddress]{revtex4-1}
\usepackage[T1]{fontenc}
\usepackage[latin9]{inputenc}
\usepackage{color}
\usepackage{babel}
\usepackage{amsmath}
\usepackage{amsthm}
\usepackage{graphicx}
\usepackage[unicode=true,pdfusetitle,
 bookmarks=true,bookmarksnumbered=false,bookmarksopen=false,
 breaklinks=false,pdfborder={0 0 0},backref=false,colorlinks=true]
 {hyperref}

\makeatletter
\theoremstyle{plain}
\newtheorem{thm}{\protect\theoremname}
  \theoremstyle{plain}
  \newtheorem{prop}[thm]{\protect\propositionname}

\usepackage{babel}
\usepackage{braket}
\usepackage{bbm}
\usepackage{txfonts}

\providecommand{\propositionname}{Proposition}
\providecommand{\theoremname}{Theorem}

\makeatother

  \providecommand{\propositionname}{Proposition}
\providecommand{\theoremname}{Theorem}

\begin{document}

\title{Structure of the resource theory of quantum coherence}

\author{Alexander Streltsov}

\email{streltsov.physics@gmail.com}

\selectlanguage{english}%

\affiliation{Faculty of Applied Physics and Mathematics, Gda\'{n}sk University
of Technology, 80-233 Gda\'{n}sk, Poland}

\affiliation{National Quantum Information Centre in Gda\'{n}sk, 81-824 Sopot,
Poland}

\affiliation{Dahlem Center for Complex Quantum Systems, Freie Universität Berlin,
D-14195 Berlin, Germany}

\author{Swapan Rana}

\affiliation{ICFO -- Institut de Ciències Fotòniques, The Barcelona Institute
of Science and Technology, 08860 Castelldefels, Spain}

\author{Paul Boes}

\author{Jens Eisert}

\affiliation{Dahlem Center for Complex Quantum Systems, Freie Universität Berlin,
D-14195 Berlin, Germany}
\begin{abstract}
Quantum coherence is an essential feature of quantum mechanics which
is responsible for the departure between classical and quantum world.
The recently established resource theory of quantum coherence studies
possible quantum technological applications of quantum coherence,
and limitations which arise if one is lacking the ability to establish
superpositions. An important open problem in this context is a simple
characterization for incoherent operations, constituted by all possible
transformations allowed within the resource theory of coherence. In
this work, we contribute to such a characterization by proving several
upper bounds on the maximum number of incoherent Kraus operators in
a general incoherent operation. For a single qubit, we show that the
number of incoherent Kraus operators is not more than 5, and it remains
an open question if this number can be reduced to 4. The presented results are also relevant for quantum thermodynamics, as we demonstrate by introducing the class of Gibbs-preserving strictly incoherent operations, and solving the corresponding mixed-state conversion problem for a single qubit.
\end{abstract}
\maketitle
Quantum resource theories \cite{Horodecki2013,Coecke2016} provide
a strong framework for studying fundamental properties of quantum
systems and their applications for quantum technology. The basis of
any quantum resource theory is the definition of \emph{free states}
and \emph{free operations}. Free states are quantum states which can
be prepared at no additional cost, while free operations capture those
physical transformations which can be implemented without consumption
of resources. Having identified these two main features, one can study
the basic properties of the corresponding theory, such as possibility
of state conversion, resource distillation, and quantification. An
important example is the resource theory of entanglement, where free
states are separable states, and free operations are local operations
and classical communication \cite{Vedral1997,Horodecki2009}.

In the resource theory of \emph{quantum coherence} \cite{Aberg2006,Baumgratz2014,Winter2016,Streltsov2016,BenDana2017},
free states are identified as incoherent states 
\begin{equation}
\rho=\sum_{i}p_{i}\ket{i}\!\bra{i},
\end{equation}
i.e., states which are diagonal in a fixed specified basis $\{\ket{i}\}$.
The choice of this basis depends on the particular problem under study,
and in many relevant scenarios such a basis is naturally singled out
by the unavoidable decoherence \cite{Zurek2003}.

The definition of free operations within the theory of coherence is
not unique, and several approaches have been discussed in the literature,
based on different physical (or mathematical) considerations \cite{Streltsov2016}.
Two important frameworks are known as incoherent \cite{Baumgratz2014}
and strictly incoherent operations \cite{Winter2016,Yadin2016}, which
will be denoted by IO and SIO, respectively. The characterizing feature
of IO is the fact that they admit an \emph{incoherent Kraus decomposition},
i.e., they can be written as \cite{Baumgratz2014} 
\begin{equation}
\Lambda(\rho)=\sum_{j}K_{j}\rho K_{j}^{\dagger},\label{eq:Lambda}
\end{equation}
where each of the Kraus operators $K_{j}$ cannot create coherence
individually, $K_{j}\ket{m}\sim\ket{n}$ for suitable integers $n$
and $m$. This approach is motivated by the fact that any quantum
operation can be interpreted as a selective measurement in which outcome
$j$ occurs with probability $p_{j}=\mathrm{Tr}[K_{j}\rho K_{j}^{\dagger}]$,
and the state after the measurement is given by $K_{j}\rho K_{j}^{\dagger}/p_{j}$.
An IO can then be interpreted as a measurement which cannot create
coherence even if one applies post-selection on the measurement outcomes.

\emph{Strictly incoherent operations} are incoherent operations with
the additional property that all $K_{i}^{\dagger}$ are also incoherent
\cite{Winter2016,Yadin2016}. These operations have several desirable
properties which distinguish them from the larger class IO. In particular,
it has been shown that SIO is the most general class of operations
which do not use coherence \cite{Yadin2016}.

Other important frameworks which are discussed in the recent literature
include maximally incoherent operations (MIO) \cite{Aberg2006}: this
is the most general class of operations which cannot create coherence
from incoherent states. It has recently been shown that this framework
has maximally coherent mixed states, i.e., quantum states which are
the optimal resource among all states with a given spectrum \cite{Streltsov2016b}.
Another important class are translationally invariant operations (TIO)
\cite{Gour2008}, these are quantum operations which commute with
time translations $e^{-iHt}$ induced by a given Hamiltonian $H$.
The set IO is strictly larger than TIO for nondegenerate Hamiltonians
\cite{Marvian2016}. Moreover, the class TIO has found several applications
in the literature, including the resource theory of asymmetry \cite{Gour2008,Marvian2016,Vaccaro2008}
and quantum thermodynamics \cite{Lostaglio2015a,Lostaglio2015b}.
Further approaches, also going beyond incoherent states, have been
investigated recently in Refs. \cite{Chitambar2016b,Chitambar2016c,Chitambar2017,Killoran2016,Theurer2017,Regula2017}. 

The quantification of coherence is another important research direction.
Postulates for coherence quantifiers have been presented \cite{Aberg2006,Baumgratz2014,Levi2014,Streltsov2016},
based on earlier approaches in entanglement theory \cite{Vedral1997,Plenio2007,Horodecki2009}.
Operational coherence measures include distillable coherence and coherence
cost \cite{Yuan2015,Winter2016}, which quantify optimal rates for
asymptotic coherence distillation and dilution via the corresponding
set of free operations. Another operational quantifier is the robustness
of coherence \cite{Napoli2016,Piani2016}, which is also closely related
to coherence quantifiers based on interferometric visibility \cite{Biswas2017}.
Distance-based coherence quantifiers have also been investigated \cite{Baumgratz2014,Streltsov2015,Rana2016},
and it was shown that the relative entropy of coherence is equal to
the distillable coherence for the sets IO and MIO \cite{Winter2016}.
For the latter set, distillable coherence coincides with the coherence
cost, which implies that the resource theory of coherence based on
MIO is reversible \cite{Winter2016}. Dynamics of coherence quantifiers
under noisy evolution has also been investigated \cite{Bromley2015,Silva2016,Mani2015,Garcia-Diaz2016,Puchala2016,Zanardi2016}.

While initially formulated for a single particle, the framework of
coherence has recently been extended to distributed scenarios \cite{Chitambar2016,Chitambar2016d,Matera2016,Streltsov2017}.
This extension has found several applications in remote quantum protocols,
including the tasks of quantum state merging \cite{Streltsov2016c,Horodecki2005}
and assisted coherence distillation \cite{Chitambar2016,Streltsov2017},
which has also been demonstrated experimentally very recently \cite{Wu2017}.
Coherence in multipartite systems has also been studied with respect
to other types of nonclassicality such as entanglement \cite{Streltsov2015,Killoran2016,Zhu2017,Regula2017}
and quantum discord \cite{Ma2016,Modi2012}.

Having identified relevant classes of free operations, it is now important
to ask about a description of those operation efficient in the physical
dimension. Such an efficient description seems crucial for a rigorous
investigation of the corresponding resource theory. In entanglement
theory, it is known that the set of local operations and classical
communication is notoriously difficult to capture mathematically \cite{Chitambar2014}.
In this work, we address this question for the resource theory of
coherence, focusing on the classes IO and SIO. We show that both classes
admit a minimal standard form, and use these results to give a full
solution for the mixed-state conversion problem via SIO, IO, and MIO. We further introduce the set of Gibbs-preserving SIO, and solve the mixed-state conversion problem for a single qubit also for these operations.

\medskip{}

\textbf{\emph{Summary of results.}} A general quantum operation, acting
on a Hilbert space of dimension $d$, admits a decomposition with
at most $d^{2}$ Kraus operators \cite{Nielsen2010} -- this is the
maximum \emph{Kraus rank}. However, since (strictly) incoherent Kraus
operators have a very specific structure, it is unclear if this result
also transfers to IO and SIO \footnote{Clearly, any IO and SIO admits a decomposition into (at most) $d^2$ Kraus operators, where $d$ is the dimension of the Hilbert space. However, it is unclear if such a minimal decomposition will have the desired (strictly) incoherent structure.}. In the following two statements, we
provide upper bounds for the number of (strictly) incoherent Kraus
operators for these operations. We refer to this minimal number as
the \emph{(strictly) incoherent Kraus rank}. 
\begin{thm}[Maximum number of incoherent Kraus operators for IO]
\label{thm:1}Any incoherent operation acting on a Hilbert space
of dimension $d$ admits a decomposition with at most $d^{4}+1$ incoherent
Kraus operators. For $d=2$ and $d=3$, this number can be improved
to 5 and 39 respectively. 
\end{thm}
\noindent This theorem is a combination of Propositions~\ref{prop:finite},
\ref{prop:qubit}, and \ref{prop:general}, which will be presented
and discussed below. The corresponding bound for SIO is given in the
following statement. 
\begin{thm}[Maximum number of strictly incoherent Kraus operators for SIO]
\label{thm:2}Any strictly incoherent operation acting on a Hilbert
space of dimension $d$ admits a decomposition with at most $\min\{d^{4}+1,\sum_{k=1}^{d}{d!}/{(k-1)!}\}$
strictly incoherent Kraus operators. 
\end{thm}
\noindent This theorem follows by combining Propositions~\ref{prop:finite}
and \ref{prop:SIO}. In general, it remains an open question if the
bounds in Theorems~\ref{thm:1} and \ref{thm:2} are tight. However,
as we prove in Proposition~\ref{prop:SIO2}, there exist SIO which
require $d^{2}$ Kraus operators. This implies that the bound in Theorem~\ref{thm:2}
is tight for SIO on a single qubit.

The above theorems assist in a rigorous investigation of the resource
theory of coherence, since they significantly reduce the number of
free parameters for the sets IO and SIO. For a single qubit, any IO
admits a decomposition into 5 incoherent Kraus operators, and for
SIO this number further reduces to 4.

\begin{figure}
\includegraphics[width=1\columnwidth]{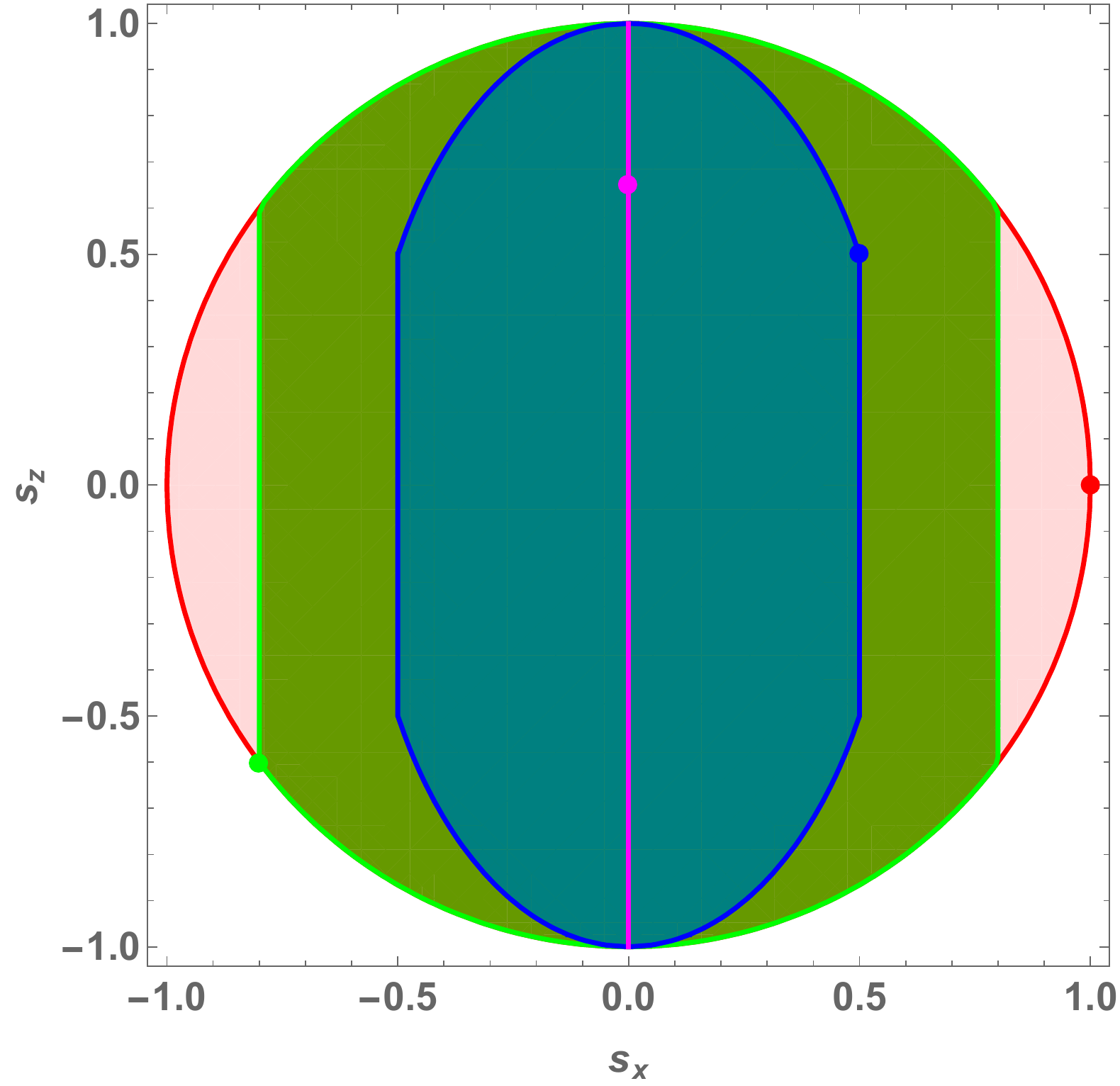}

\caption{\label{fig:SIO}Achievable region for single-qubit SIO, IO, and MIO.
Colored areas show the projection of the achievable region in the
$x$-$z$ plane for initial Bloch vectors $(0.5,0,0.5)^{T}$ {[}blue{]},
$(-0.8,0,-0.6)^{T}$ {[}green{]}, and $(1,0,0)^{T}$ {[}red{]}. Note
that the last two states are pure. The magenta line corresponds to
the achievable region of an incoherent state with Bloch vector $(0,0,0.65)^{T}$.}
\end{figure}
We will now demonstrate the power of these results by providing a
full characterization for all possible state transformations via single-qubit
SIO, IO, and MIO. As we show in Appendix~\ref{sec:Reachable-states},
a single-qubit state $\rho$ with Bloch vector $(r_{x},r_{y},r_{z})^{T}$
can be converted into another single-qubit state $\sigma$ with Bloch
vector $(s_{x},s_{y},s_{z})^{T}$ via SIO, IO, or MIO if and only
if the following inequalities are fulfilled: \begin{subequations}\label{eq:region}
\begin{eqnarray}
s_{x}^{2}+s_{y}^{2} & \leq & r_{x}^{2}+r_{y}^{2},\label{eq:region-1}\\
s_{z}^{2} & \leq & 1-\frac{1-r_{z}^{2}}{r_{x}^{2}+r_{y}^{2}}(s_{x}^{2}+s_{y}^{2}).\label{eq:region-2}
\end{eqnarray}
\end{subequations}

\noindent For a given initial Bloch vector $(r_{x},r_{y},r_{z})^{T}$,
these inequalities completely characterize the achievable region for
the final Bloch vectors $(s_{x},s_{y},s_{z})^{T}$. The achievable
region is symmetric under rotations around the $z$-axis and corresponds
to a cylinder with radius $(r_{x}^{2}+r_{y}^{2})^{1/2}$ and height
$2r_{z}$ with ellipsoids attached at the top and the bottom. In Fig.~\ref{fig:SIO}
we show the projection of the achievable region into the $x$-$z$
plane for four initial states. The proof of Eqs.~(\ref{eq:region})
in Appendix~\ref{sec:Reachable-states} makes use of our result that
any single-qubit SIO can be decomposed into four strictly incoherent
Kraus operators, see Theorem~\ref{thm:2} and discussion below Proposition~\ref{prop:qubit}
for the general form of these operators. Alternatively, the form
of the achievable region can be proven using results in \cite{Chitambar2016b,Chitambar2016c,Chitambar2017}.
We also note that for pure states the conversion problem has been
completely solved for SIO \cite{Winter2016} and IO \cite{Zhu2017}.

\begin{figure}
\includegraphics[width=1\columnwidth]{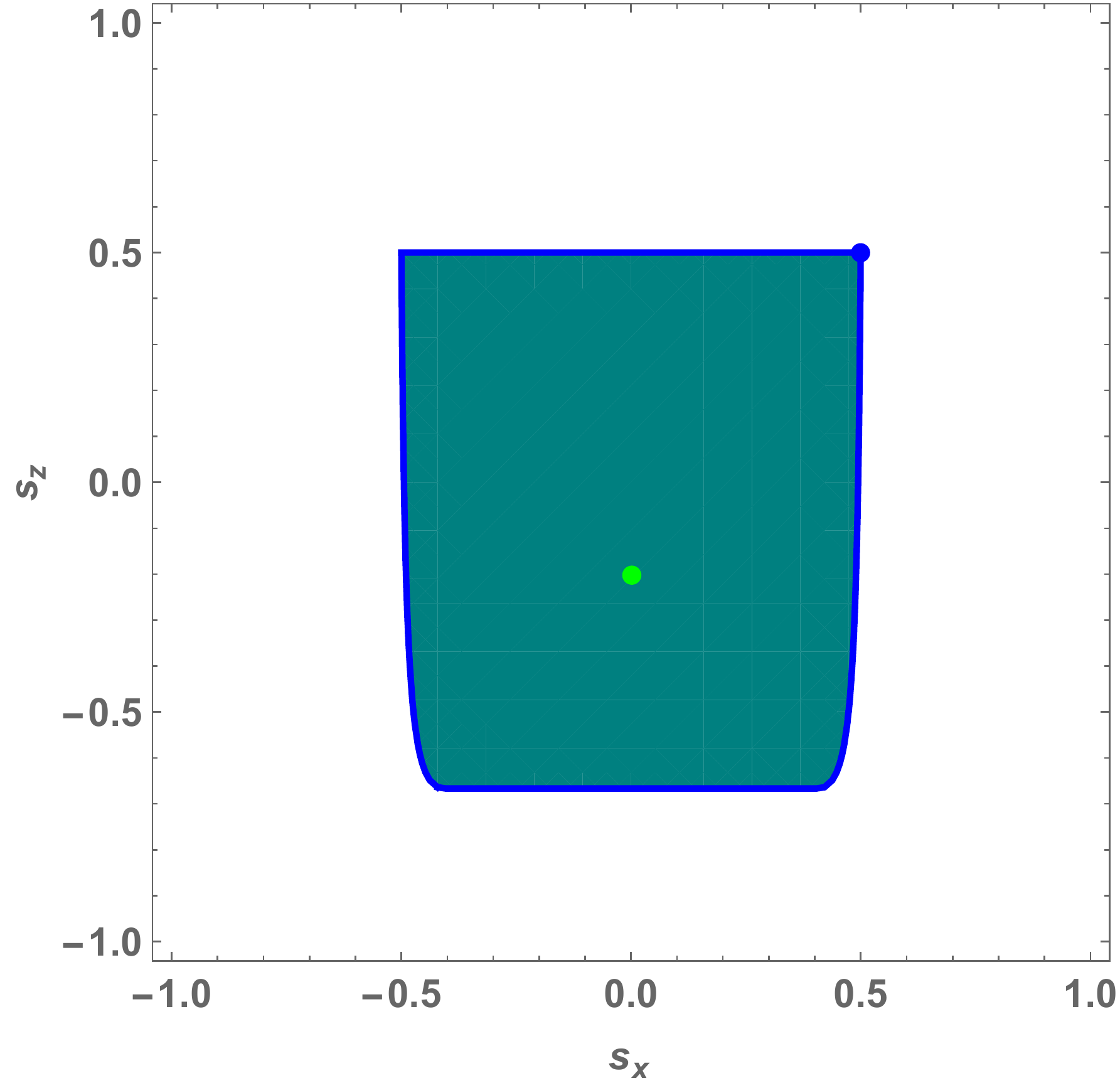}

\caption{\label{fig:2}Achievable region {[}blue area{]} of Gibbs-preserving
SIO on a single qubit. The initial state has the Bloch vector $\boldsymbol{r}=(0.5,0,0.5)^{T}$
{[}blue dot{]} and the preserved Gibbs state has the Bloch vector
$\boldsymbol{t}=(0,0,-0.2)^{T}$ {[}green dot{]}.}

\end{figure}

As a second application for our results, we investigate strictly incoherent
operations which preserve a given incoherent state $\tau$, i.e.,
\begin{equation}
\Lambda(\tau)=\tau.\label{eq:Gibbs}
\end{equation}
The motivation for this constraint originates from quantum thermodynamics.
In particular, if $\tau=e^{-\beta H}/\mathrm{Tr}[e^{-\beta H}]$ is
the Gibbs state of a system with Hamiltonian $H$, then the condition~(\ref{eq:Gibbs})
is known to hold for thermal operations \cite{Janzing2000,Lostaglio2015b,Lostaglio2015a}.
For a non-degenerate Hamitonian $H$ thermal operations cannot create
coherence in the eigenbasis of $H$, and conditions for state transformations
under these operations and the role of coherence therein have been
extensively studied in Refs.\ \cite{Horodecki2013b,Brandao2015,Cwiklinski2015}.
The most general quantum operations that fulfill Eq.~(\ref{eq:Gibbs})
are known as Gibbs-preserving operations, and it has been shown that
such operations can create coherence \cite{Faist2015}. Here, we contribute
to this discussion by introducing the set of \emph{Gibbs-preserving SIO}, and giving a full solution for the mixed-state conversion problem on a single qubit. As an example,
in Fig.~\ref{fig:2} we show the achievable region in the $x$-$z$
plane for an initial state $\rho$ with Bloch vector $\boldsymbol{r}=(0.5,0,0.5)^{T}$,
and the preserved Gibbs state $\tau$ has the Bloch vector $\boldsymbol{t}=(0,0,-0.2)^{T}$.
Note that the achievable region is convex and symmetric under rotations
around the $z$-axis. We refer to Appendix~\ref{sec:B} for more
details and further examples.

This discussion clearly demonstrates how Theorems~\ref{thm:1}
and \ref{thm:2} lead to deep insights on the structure of the resource
theory of quantum coherence. In particular, they lead to a full solution
of the state conversion problem under single-qubit SIO, IO, and MIO.
These results have clear relevance within the resource theory of coherence,
and also extend to other related fields, including quantum thermodynamics.

In the remainder of this letter we will present statements which are
needed for proving the aforementioned theorems, and provide further
detailed discussion of the results.

\medskip{}

\textbf{\emph{Bounds from Choi-Jamio\l kowski isomorphism.}} We will
now present bounds for IO and SIO arising from the \emph{Choi-Jamio\l kowski
isomorphism} between a quantum operation $\Lambda$ and the corresponding
Choi state 
\begin{equation}
\rho_{\Lambda}=(\Lambda\otimes\mathrm{id})(\Phi_{d}^{+}),
\end{equation}
where $\Phi_{d}^{+}=d^{-1}\sum_{i,j=0}^{d-1}\ket{i,i}\!\bra{j,j}$
is a maximally entangled state of dimension $d^{2}$. The rank of
the Choi state is the \emph{Kraus rank}, which is the smallest number
of (not necessarily incoherent) Kraus operators.
\begin{prop}[Bounds originating from the Choi state]
\label{prop:finite}Any (strictly) incoherent operation acting on
a Hilbert space of dimension $d$ admits a decomposition with at most
$d^{4}+1$ (strictly) incoherent Kraus operators.
\end{prop}
\noindent The proof can be found in Appendix~\ref{sec:Proof-Finite}.
It is interesting to note that the proof of the above proposition
does not used the fact that the Kraus operators are incoherent. Thus,
the proposition is not limited to IO, but also applies to SIO. Moreover,
it also applies to separable operations, i.e., operations of the form
\begin{equation}
\Lambda_{\mathrm{s}}(\rho)=\sum_{i}\left(A_{i}\otimes B_{i}\right)\rho(A_{i}^{\dagger}\otimes B_{i}^{\dagger}).\label{eq:separable_operation}
\end{equation}
Any separable operation can be decomposed into (at most) $d^{4}+1$
product Kraus operators $A_{i}\otimes B_{i}$, where $d=d_{A}d_{B}$
is the dimension of the total system. The same is true for separable
quantum-incoherent operations \cite{Streltsov2017}, i.e., operations
of the form~(\ref{eq:separable_operation}) with incoherent operators
$B_{i}$, and separable incoherent operations where both $A_{i}$
and $B_{i}$ are incoherent \cite{Streltsov2017}. For single-qubit
IO and SIO the upper bound in Proposition~\ref{prop:finite} gives
17 operators. As we will show in the following, this number can be
significantly reduced.

\medskip{}

\textbf{\emph{Bounds from the structure of (strictly) incoherent operations.
}}We will now provide improved bounds which explicitly make use of
the structure of IO and SIO. For this, we will first consider incoherent
operations on a single-qubit, i.e., the corresponding Hilbert space
has dimension $d=2$. The following proposition shows that any single-qubit
IO can be decomposed into 5 incoherent Kraus operators with a simple
structure.
\begin{prop}[Incoherent operations on qubits]
\label{prop:qubit} Any incoherent operation acting on a single qubit
admits a decomposition with at most 5 incoherent Kraus operators.
A canonical choice of the operators is given by the set 
\begin{equation}
\left\{ \begin{pmatrix}a_{1} & b_{1}\\
0 & 0
\end{pmatrix},\begin{pmatrix}0 & 0\\
a_{2} & b_{2}
\end{pmatrix},\begin{pmatrix}a_{3} & 0\\
0 & b_{3}
\end{pmatrix},\begin{pmatrix}0 & b_{4}\\
a_{4} & 0
\end{pmatrix},\begin{pmatrix}a_{5} & 0\\
0 & 0
\end{pmatrix}\right\} ,\label{eq:IO-qubit}
\end{equation}
where $a_{i}$ can be chosen real, while $b_{i}\in\mathbbm{C}$. Moreover,
it holds that $\sum_{i=1}^{5}a_{i}^{2}=\sum_{j=1}^{4}|b_{j}|^{2}=1$
and $a_{1}b_{1}+a_{2}b_{2}=0$. 
\end{prop}
\noindent We refer to Appendix~\ref{sec:Proof-Qubit} for the proof.
We also note that the same techniques can be applied to single-qubit
SIO, in which case the number of strictly incoherent operators reduces
to 4. The corresponding general form of strictly incoherent Kraus
operators can be given as 
\begin{equation}
\left\{ \begin{pmatrix}a_{1} & 0\\
0 & b_{1}
\end{pmatrix},\begin{pmatrix}0 & b_{2}\\
a_{2} & 0
\end{pmatrix},\begin{pmatrix}a_{3} & 0\\
0 & 0
\end{pmatrix},\begin{pmatrix}0 & 0\\
a_{4} & 0
\end{pmatrix}\right\} ,
\end{equation}
where $a_{i}$ are real and $\sum_{i=1}^{4}a_{i}^{2}=\sum_{j=1}^{2}|b_{j}|^{2}=1$.
A more general bound for SIO for arbitrary dimensions will be given
below.

It is important to note that the proofs of Propositions \ref{prop:finite}
and \ref{prop:qubit} are fundamentally different: while the argument
leading to Proposition \ref{prop:finite} is based on the Choi-Jamio\l kowski
isomorphism and Caratheodory's theorem, the proof of Proposition~\ref{prop:qubit}
makes explicit use of the structure of IO. In the next step, we will
extend Proposition \ref{prop:qubit} to arbitrary dimension.
\begin{prop}[IO for $d$-dimensional systems]
\label{prop:general}Any incoherent operation for a quantum systems
of dimension $d$ admits a decomposition with at most $d(d^{d}-1)/(d-1)$
incoherent Kraus operators. 
\end{prop}
\noindent We refer to Appendix~\ref{sec:Proof-General} for the proof.
For single-qubit IO Proposition \ref{prop:general} gives us 6 incoherent
operators as an upper bound. As we have already seen in Proposition~\ref{prop:qubit},
this number can be reduced to 5. For qutrits we obtain 39 Kraus operators,
while the bound in Proposition~\ref{prop:finite} gives 82 Kraus
operators. For dimensions larger than 3 Proposition~\ref{prop:finite}
always gives a better bound.

We will now see how the tools presented above can be applied to study
the structure of SIO. By Proposition~\ref{prop:finite}, any SIO
admits a decomposition into (at most) $d^{4}+1$ strictly incoherent
Kraus operators. As we will show in the following, for small dimensions
this number can be significantly reduced. 
\begin{prop}[SIO for $d$-dimensional systems]
\label{prop:SIO}Any strictly incoherent operation acting on a Hilbert
space of dimension $d$ admits a decomposition with at most $\sum_{k=1}^{d}{d!}/{(k-1)!}$
strictly incoherent Kraus operators. 
\end{prop}
\noindent The proof of the proposition can be found in Appendix~\ref{sec:Proof-prop:SIO}.
Note that the bound in this proposition is below $d^{4}+1$ for $d\leq5$.
For larger dimensions $d^{4}+1$ gives a better bound. For $d=2$
we see that any single-qubit SIO admits a decomposition into 4 strictly
incoherent Kraus operators. This was already discussed below Proposition~\ref{prop:qubit}.
As we will show in the following, this bound is tight.
\begin{prop}[Lower bound]
\label{prop:SIO2}For a Hilbert space of dimension $d$, there exist
strictly incoherent operations which cannot be implemented with fewer
than $d^{2}$ Kraus operators.
\end{prop}
\noindent The proof of the proposition can be found in Appendix~\ref{sec:Lower-Bound}.
Note that for $d=2$ the bounds in Propositions \ref{prop:SIO} and
\ref{prop:SIO2} coincide: any single-qubit SIO can be decomposed
into 4 strictly incoherent Kraus operators, and some single-qubit
SIO require 4 Kraus operators in their decomposition.

\medskip{}

\textbf{\emph{Conclusions.}} In this work we have studied the structure
of the resource theory of quantum coherence, focusing in particular
on the structure of incoherent and strictly incoherent operations.
We have shown that any (strictly) incoherent operation can be written
with at most $d^{4}+1$ (strictly) incoherent Kraus operators, where
$d$ is the dimension of the Hilbert space under study. For small
dimensions this number can be significantly reduced. For a single
qubit any IO can be decomposed into 5 incoherent Kraus operators,
while any SIO admits a decomposition into 4 strictly incoherent Kraus
operators. While the latter bound is tight, the tightness of the other
bounds remains an open question. 

Our results assist in the systematic investigation of the resource
theory of coherence due to the significant reduction of unknown parameters.
We have applied our results to solve the mixed-state conversion
problem for single-qubit SIO, IO, and MIO. We further introduced the
class of Gibbs-preserving strictly incoherent operations and also
solved the corresponding mixed-state conversion problem for a single
qubit. A natural next step would be to consider single-qubit incoherent
operations applied on one subsystem of a bipartite quantum state.
Such multipartite scenarios have been previously studied in Refs.\ \cite{Chitambar2016,Chitambar2016d,Matera2016,Streltsov2017},
and the results presented in this letter provide a strong framework
for their further investigation. Another important question which
is left open in this work is the relation of Gibbs-preserving strictly
incoherent operations to thermal operations. We expect that further
results in this direction will be presented in the near future, exploring Gibbs-preserving strictly incoherent operations in relation to recent works on quantum thermodynamics \cite{Lostaglio2015a,Lostaglio2015b,Horodecki2013b,Brandao2015,Cwiklinski2015} and extending them to other notions of quantum coherence \cite{Streltsov2016}.

Finally, our results also transfer to other related concepts, including
translationally invariant operations, which are relevant in the resource
theory of asymmetry and quantum thermodynamics. Recalling that TIO
is a subset of IO, the results presented in this letter also give
bounds on decompositions of TIO into incoherent Kraus operators. Thus,
numerical simulations and optimizations over all these classes now
also become feasible, at least for small dimensions.

\textbf{\emph{Note added.}} Upon completion of this manuscript, a
related work has appeared on the arXiv \cite{Shi2017}.

\textbf{\emph{Acknowledgements.}} We thank Gerardo Adesso, Dario Egloff,
Martin Plenio, Thomas Theurer, Henrik Wilming, and especially Andreas
Winter for discussions. We acknowledge financial support by the Alexander
von Humboldt-Foundation, the National Science Center in Poland (POLONEZ
UMO-2016/21/P/ST2/04054), EU grants OSYRIS (ERC-2013-AdG Grant No.\ 339106),
TAQ (ERC CoG Grant No.\ 307498) and QUIC (H2020-FETPROACT-2014 Grant
No. 641122), the Spanish MINECO grant FISICATEAMO (FIS2016-79508-P),
the Severo Ochoa Programme (SEV-2015-0522), MINECO CLUSTER (ICFO15-EE-3785),
the Generalitat de Catalunya (2014 SGR 874 and CERCA/Program), the
Fundació Privada Cellex, the DFG (CRC 183), and Studienstiftung des
Deutschen Volkes.

\bibliographystyle{apsrev4-1}
\bibliography{literature}

\appendix

\section{\label{sec:Reachable-states}Proof of Eqs.~(\ref{eq:region-1})
and (\ref{eq:region-2})}

In the following, we will prove that via SIO it is possible to convert
a single-qubit state $\rho$ with Bloch vector $\boldsymbol{r}=(r_{x},r_{y},r_{z})^{T}$
into another single-qubit state $\sigma$ with Bloch vector $\boldsymbol{s}=(s_{x},s_{y},s_{z})^{T}$
if and only if 
\begin{eqnarray}
s_{x}^{2}+s_{y}^{2} & \leq & r_{x}^{2}+r_{y}^{2},\\
s_{z}^{2} & \leq & 1-\frac{1-r_{z}^{2}}{r_{x}^{2}+r_{y}^{2}}(s_{x}^{2}+s_{y}^{2}).
\end{eqnarray}
Note that by Theorem 3 in Ref.\ \cite{Chitambar2016b}, this also proves
that these conditions are necessary and sufficient for state transformations
via single-qubit IO and MIO.

In the first step, we will discuss the symmetries and main properties
of the considered problem. For a given initial state $\rho$, the
achievable region (i.e., the set of all states which is achievable
via SIO) is invariant under rotations around the $z$-axis of the
Bloch sphere. This follows from the fact that such rotations are strictly
incoherent unitaries. The achievable region is further symmetric under
reflection at the $x$-$y$ plane, since such a reflection corresponds
to a $\sigma_{x}$ rotation, which is SIO. Moreover, the achievable
region is convex and contains the maximally mixed state $\openone/2$.

Due to these properties we can significantly simplify our further
analysis by focusing on the positive part of the $x$-$z$ plane,
i.e., we set $r_{y}=s_{y}=0$, and all the other coordinates are nonnegative.
Our statement then reduces to the following inequalities: 
\begin{eqnarray}
s_{x}^{2} & \leq & r_{x}^{2},\label{eq:Region-x}\\
s_{z}^{2} & \leq & 1-\frac{1-r_{z}^{2}}{r_{x}^{2}}s_{x}^{2}.\label{eq:Region-z}
\end{eqnarray}
In a next step, we note that any single-qubit SIO can be decomposed
into 4 Kraus operators of the form 
\begin{equation}
\begin{aligned}K_{1} & =\left(\begin{array}{cc}
a_{1} & 0\\
0 & b_{1}
\end{array}\right),\,\,\,\,K_{2}=\left(\begin{array}{cc}
0 & b_{2}\\
a_{2} & 0
\end{array}\right),\\
K_{3} & =\left(\begin{array}{cc}
a_{3} & 0\\
0 & 0
\end{array}\right),\,\,\,\,K_{4}=\left(\begin{array}{cc}
0 & b_{3}\\
0 & 0
\end{array}\right).
\end{aligned}
\label{eq:SIOkraus}
\end{equation}
In particular, any SIO is completely determined by the vectors $\boldsymbol{a}=(a_{1},a_{2},a_{3})^{T}$
and $\boldsymbol{b}=(b_{1},b_{2},b_{3})^{T}$. The normalization condition
$\sum_{i=1}^{4}K_{i}^{\dagger}K_{i}=\openone$ implies that $|\boldsymbol{a}|=|\boldsymbol{b}|=1$.
Moreover, $\boldsymbol{a}$ can be chosen real. The proof for this
form of single-qubit SIOs follows the same line of reasoning as the
proof of Proposition~\ref{prop:qubit}, see also Appendix~\ref{sec:Proof-Qubit}
for more details. By using the explicit form of the Kraus operators
(\ref{eq:SIOkraus}), the Bloch coordinates $s_{x}$ and $s_{z}$
of the final state take the form \begin{subequations} \label{eq:SxSz}
\begin{eqnarray}
s_{x} & = & r_{x}\left(a_{1}\mathrm{Re}\left[b_{1}\right]+a_{2}\mathrm{Re}\left[b_{2}\right]\right),\label{eq:Sx}\\
s_{z} & = & 1-a_{2}^{2}(1+r_{z})-|b_{1}|^{2}(1-r_{z}),\label{eq:Sz}
\end{eqnarray}
\end{subequations} where $\mathrm{Re}[b_{i}]=(b_{i}+b_{i}^{*})/2$
denotes the real part of the complex number $b_{i}$.

We will now focus on the boundary of the achievable region in the
positive part of the $x$-$z$ plane. In particular, we will show
that this boundary can be attained with real vectors $\boldsymbol{a}$
and $\boldsymbol{b}$ with $a_{3}=b_{3}=0$. For this, let $(s_{x},s_{z})$
be a point at the boundary. Assume now that the corresponding vectors
$\boldsymbol{a}$ and $\boldsymbol{b}$ do not have the aforementioned
properties. Then, we can introduce the following normalized real vectors
\begin{eqnarray}
\boldsymbol{a}' & = & ([a_{1}^{2}+a_{3}^{2}]^{1/2},|a_{2}|,0)^{T},\\
\boldsymbol{b}' & = & (|b_{1}|,[|b_{2}|^{2}+|b_{3}|^{2}]^{1/2},0)^{T}.
\end{eqnarray}
The vectors $\boldsymbol{a}'$ and $\boldsymbol{b}'$ give rise to
a state with Bloch coordinates $s_{x}'$ and $s_{z}'$, as given in
Eqs.~(\ref{eq:Sx}) and (\ref{eq:Sz}). It is now straightforward
to check that $s_{z}'=s_{z}$ and $s_{x}'\geq s_{x}$. However, we
assumed that $(s_{x},s_{z})$ belongs to the boundary. Since the achievable
region is convex, it must be that $s_{x}'=s_{x}$. This proves that
the boundary can be attained with real vectors $\boldsymbol{a}$ and
$\boldsymbol{b}$ with $a_{3}=b_{3}=0$.

In the final step, we introduce the parametrization 
\begin{eqnarray}
a_{1} & = & \cos\frac{\theta-\phi}{2},\,\,\,a_{2}=\sin\frac{\theta-\phi}{2},\\
b_{1} & = & \sin\frac{\theta+\phi}{2},\,\,\,b_{2}=\cos\frac{\theta+\phi}{2}.
\end{eqnarray}
Using these parameters in Eqs.~(\ref{eq:Sx}) and (\ref{eq:Sz})
we obtain 
\begin{eqnarray}
s_{x} & = & r_{x}\sin\theta,\label{eq:sx-1}\\
s_{z} & = & r_{z}\sin\theta\sin\phi+\cos\theta\cos\phi.\label{eq:sz-1}
\end{eqnarray}
Recall that due to the symmetry of the problem we are considering
only nonnegative values of $r_{x}$ and $s_{x}$. Thus, without loss
of generality, the parameter $\theta$ can be chosen from the range
$0\leq\theta\leq\pi/2$.

From Eq.~(\ref{eq:sx-1}) we can immediately verify Eq.~(\ref{eq:Region-x}),
and it remains to prove Eq.~(\ref{eq:Region-z}). For this, we will
evaluate the maximal value for $s_{z}$, achievable for given values
of $r_{x}$, $r_{z}$, and $s_{x}$. The maximum of $s_{z}$ is determined
by $ds_{z}/d\phi=0$. Taking the aforementioned parameter regions
into account, we can solve this problem explicitly for $\phi$, with
the solution 
\begin{eqnarray}
\phi & =\arctan & \left[r_{z}\tan\theta\right].
\end{eqnarray}
Using this solution in Eq.~(\ref{eq:sz-1}) we obtain the explicit
form for the maximal $s_{z}$ 
\begin{equation}
s_{z}=({\cos^{2}\theta+r_{z}^{2}\sin^{2}\theta})^{1/2}.\label{eq:sz-2}
\end{equation}

In summary, we have shown that the points $(r_{x}\sin\theta,[\cos^{2}\theta+r_{z}^{2}\sin^{2}\theta]^{1/2})$
with parameter $\theta\in[0,\pi/2]$ are located at the boundary of
the achievable region in the positive $x$-$z$ plane. Note that all
these points fulfill Eq.~(\ref{eq:Region-z}) with equality. Moreover,
this result completely characterizes the full achievable region due
to convexity, rotational symmetry around the $z$-axis, and reflection
symmetry at the $x$-$y$ plane. This completes the proof. We note
that the statement can also be proven in an alternative way, by using
results in section III of Ref.\ \cite{Chitambar2016c}.

\section{\label{sec:B}Gibbs-preserving SIO}

\begin{figure*}
\centering{}\includegraphics[width=0.95\columnwidth]{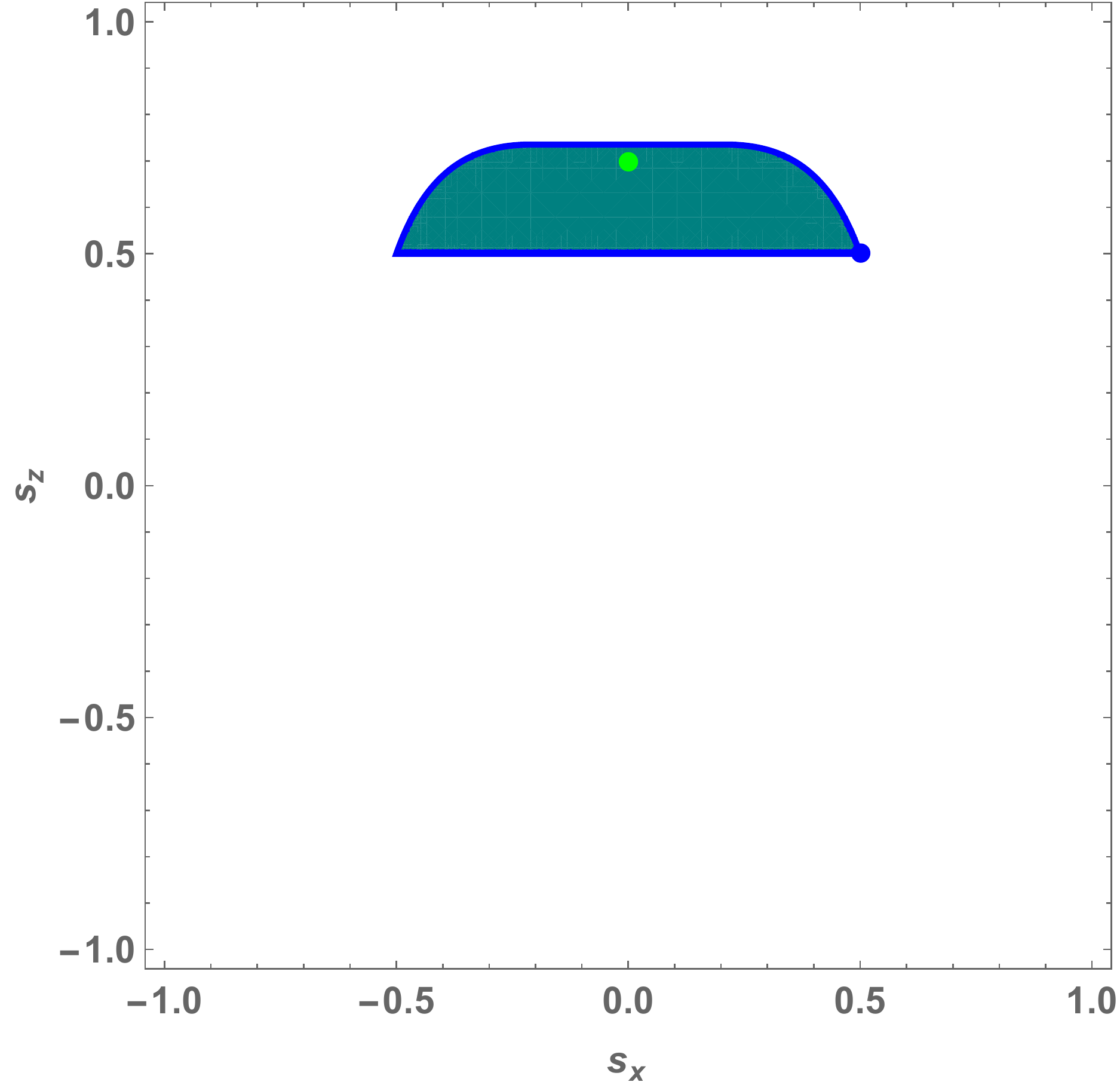}~~~~~\includegraphics[width=0.95\columnwidth]{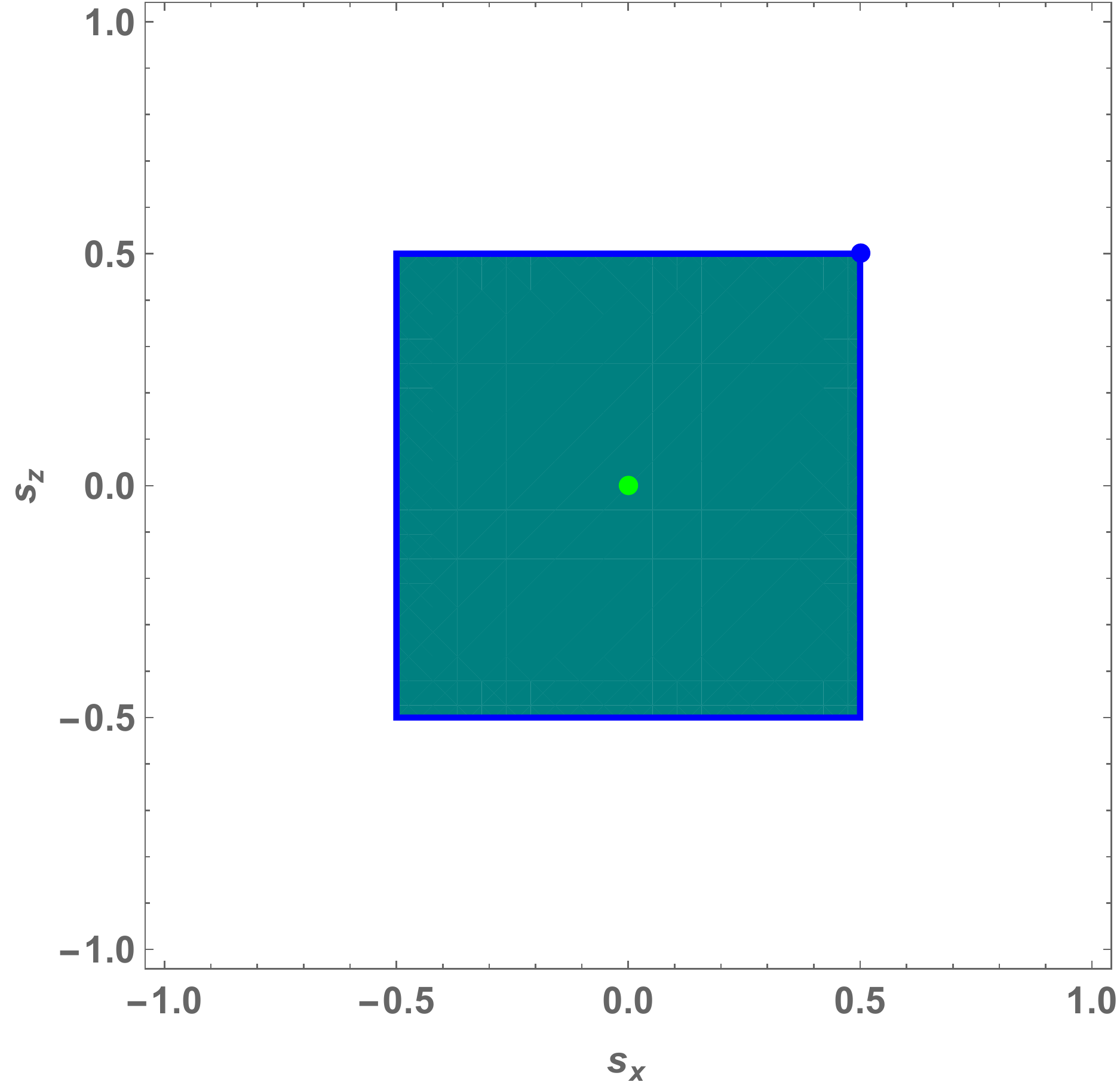}\caption{\label{fig:3}Achievable region {[}blue area{]} of single-qubit
SIO which preserve the state $\boldsymbol{t}=(0,0,0.7)^{T}$ {[}left
figure{]} and $\boldsymbol{t}=(0,0,0)^{T}$ {[}right figure{]}. The
initial state has the Bloch vector $\boldsymbol{r}=(0.5,0,0.5)^{T}$
{[}blue dot{]}, and the corresponding Bloch vector $\boldsymbol{t}$
is shown as a green dot.}
\end{figure*}
In the following, we consider single-qubit SIO which preserve a given
incoherent state $\tau$, i.e., 
\begin{equation}
\Lambda(\tau)=\tau,\label{eq:fixed_point}
\end{equation}
where the Bloch vector corresponding to $\tau$ is given by $\boldsymbol{t}=(0,0,t_{z})$. 

Before presenting the solution for this problem, we note that any single-qubit incoherent state 
\begin{equation}
\tau = p \ket{0}\!\bra{0} + (1-p) \ket{1}\!\bra{1}
\end{equation}
can be interpreted as a Gibbs state
$\tau=e^{-\beta H}/\mathrm{Tr}[e^{-\beta H}]$
for a suitable inverse temperature $\beta=\frac{1}{kT}$ and Hamiltonian $H$ which is diagonal in the incoherent basis. For $0<p<1$ and $p \neq 1/2$ this can be seen by choosing $\beta$ and $H$ such that
\begin{equation}
p = \frac{e^{-\beta E_0}}{e^{-\beta E_0} + e^{-\beta E_1}},
\end{equation}
where $E_0$ and $E_1$ are the eigenvalues of the Hamiltonian. Such choice of parameters is always possible for $0<p<1$ and $p \neq 1/2$. The cases $p=0$ and $p=1$ can be obtained by taking the limits $E_0 \rightarrow \infty$ and $E_1 \rightarrow \infty$, respectively. The case $p=1/2$ is obtained by taking the limit of infinite temperature, i.e., $\beta \rightarrow 0$.

In the following, we will determine the achievable region of these operations. The
achievable region is convex and invariant under rotations around the
$z$-axis of the Bloch sphere, as such rotations are strictly incoherent
unitaries which preserve any incoherent state. We can thus restrict
ourselves to the $x$-$z$ plane in the following discussion.

We now recall that any single-qubit SIO admits a decomposition into
four Kraus operators of the form~(\ref{eq:SIOkraus}), and the coordinates
$s_{x}$ and $s_{z}$ of the final Bloch vector take the form given
in Eqs.~(\ref{eq:SxSz}). Moreover, the condition in Eq.~(\ref{eq:fixed_point})
implies the  equality
\begin{equation}
(1+t_{z})a_{2}^{2}=(1-t_{z})(1-|b_{1}|^{2}).\label{eq:constraint}
\end{equation}
This equality leads to constraints on the range of the parameter $a_{2}$.
In particular, $a_{2}$ can take any real value compatible with 
\begin{equation}
0\leq a_{2}^{2}\leq\min\left\{ 1,\frac{1-t_{z}}{1+t_{z}}\right\} .\label{eq:a2}
\end{equation}

\begin{figure*}
\centering{}\includegraphics[width=0.95\columnwidth]{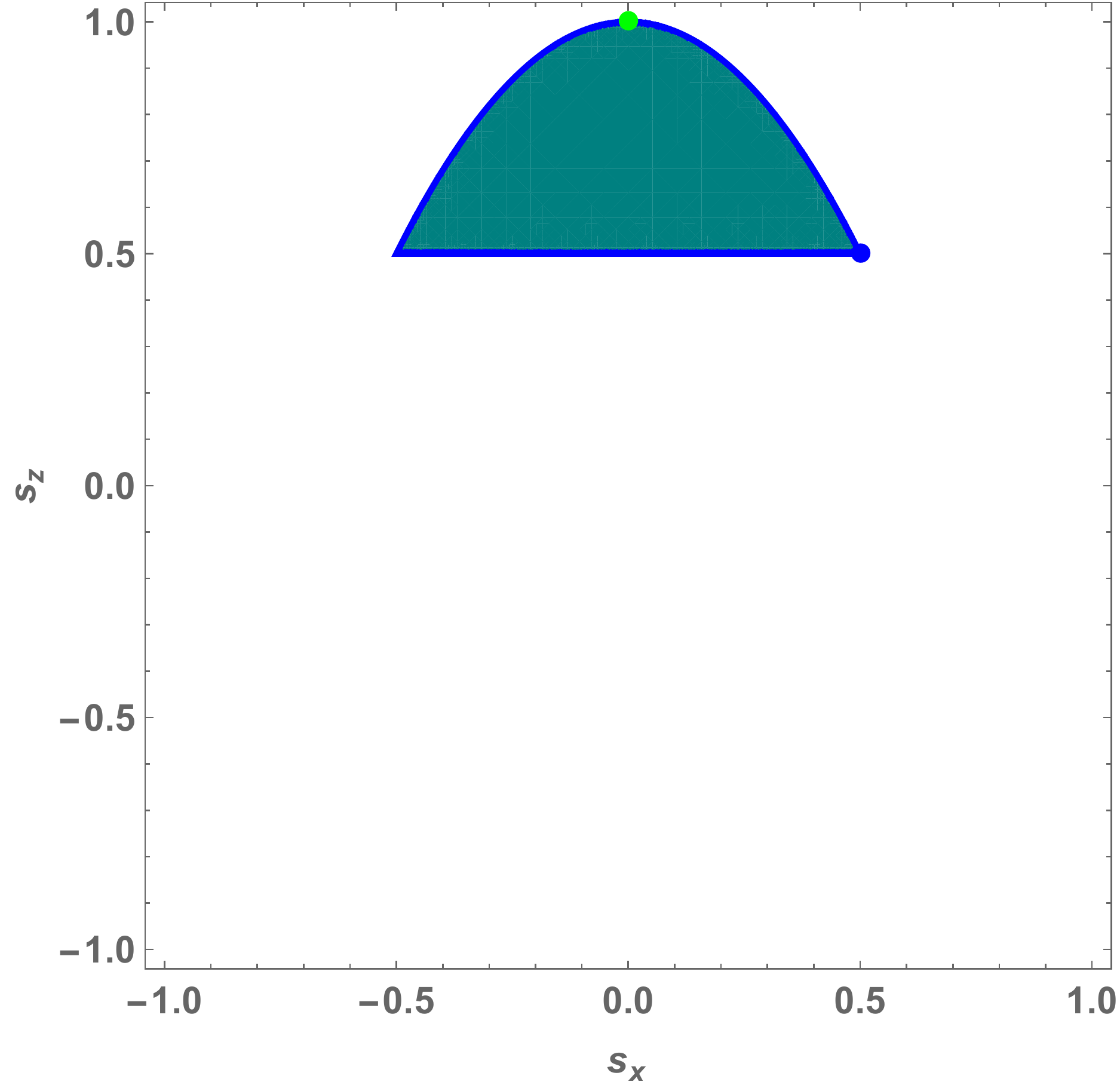}~~~~~\includegraphics[width=0.95\columnwidth]{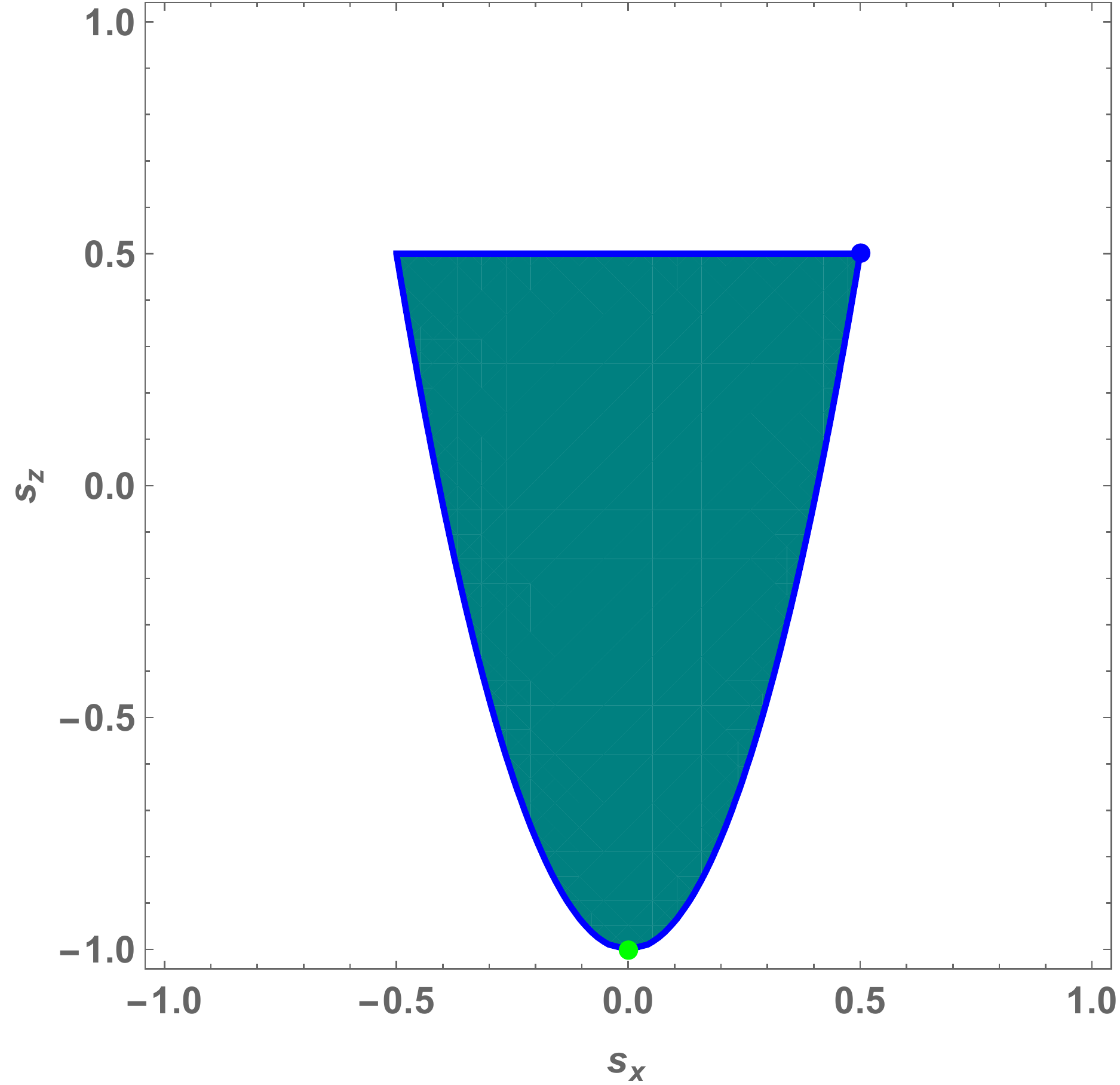}\caption{\label{fig:4}Achievable region {[}blue area{]} of single-qubit
SIO which preserve the state $\boldsymbol{t}=(0,0,1)$ {[}left figure{]}
and $\boldsymbol{t}=(0,0,-1)$ {[}right figure{]}. The initial state
{[}blue dot{]} is the same as in Fig.~\ref{fig:3}, and the corresponding
Bloch vector $\boldsymbol{t}$ is shown as a green dot.}
\end{figure*}
Before we consider the most general case, let us investigate the situation
where the SIO is unital, i.e., $t_{z}=0$. As we will now show, in
this case the achievable region in the $x$-$z$ plane is determined
by
\begin{equation}
s_{x}^{2}\leq r_{x}^{2},\,\,\,\,s_{z}^{2}\leq r_{z}^{2}.\label{eq:unital}
\end{equation}
 For proving this, note that for $t_{z}=0$ we obtain from Eq.~(\ref{eq:constraint})
\begin{equation}
|b_{1}|^{2}=1-a_{2}^{2}.
\end{equation}
Moreover, from Eq.~(\ref{eq:a2}) we see that $a_{2}$ can take any
value between $-1$ and $1$. Using these results in Eq.~(\ref{eq:Sz}),
we obtain 
\begin{align}
s_{z} & =r_{z}-2a_{2}^{2}r_{z},
\end{align}
and it follows that $s_{z}$ can take any value between $-|r_{z}|$
and $|r_{z}|$. It remains to show that $s_{x}$ can take any value
between $-|r_{x}|$ and $|r_{x}|$ for any given value of $s_{z}$.
This can be seen by choosing the parameters $a_{i}$ and $b_{i}$
as follows, 
\begin{align}
a_{1} & =b_{1}=\sqrt{1-a_{2}^{2}},\\
a_{2} & =b_{2}=\sqrt{\frac{r_{z}-s_{z}}{2r_{z}}},\\
a_{3} & =b_{3}=0.
\end{align}
For this choice of parameters Eq.~(\ref{eq:Sx}) gives us $s_{x}=r_{x}$.
Note that a larger value for $s_{x}$ cannot be achieved via SIO in
general, see also the discussion in Appendix~\ref{sec:Reachable-states}.
By symmetry, this completes the proof of Eq.~(\ref{eq:unital}).
In general (i.e. without restricting to the $x$-$z$ plane), the
achievable region for unital single-qubit SIOs is given by
\begin{equation}
s_{x}^{2}+s_{y}^{2}\leq r_{x}^{2}+r_{y}^{2},\,\,\,\,s_{z}^{2}\leq r_{z}^{2}.
\end{equation}
In the right part of Fig.~\ref{fig:3} we show this region for the
initial state with the Bloch vector $\boldsymbol{r}=(0.5,0,0.5)^{T}$.

In the next step we will focus on another special case, namely $t_{z}=1$.
From Eq.~(\ref{eq:constraint}) we immediately see that $a_{2}=0$,
and $|b_{1}|$ can take any value between 0 and 1. From Eq.~(\ref{eq:Sz})
we further obtain 
\begin{equation}
s_{z}=1-|b_{1}|^{2}(1-r_{z}),\label{eq:Sz-2}
\end{equation}
which means that $s_{z}$ can take any value in the range $r_{z}\leq s_{z}\leq1$.
We now use Eq.~(\ref{eq:Sx}), noting that for $a_{2}=0$ and fixed
value of $|b_{1}|$ the maximal value of $s_{x}$ is given as $s_{x}=|r_{x}b_{1}|$.
Together with Eq.~(\ref{eq:Sz-2}), we finally obtain 
\begin{equation}
s_{x}=\left|r_{x}\right|\sqrt{\frac{1-s_{z}}{1-r_{z}}}
\end{equation}
for the maximal value of $s_{x}$. By symmetry, the achievable region
in the $x$-$z$ plane is thus determined by the inequalities 
\begin{align}
\frac{s_{x}^{2}}{r_{x}^{2}} & \leq\frac{1-s_{z}}{1-r_{z}},\\
r_{z} & \leq s_{z}\leq1.
\end{align}
In general (i.e., without restricting to the $x$-$z$ plane) the
achievable region is given by 
\begin{align}
\frac{s_{x}^{2}+s_{y}^{2}}{r_{x}^{2}+r_{y}^{2}} & \leq\frac{1-s_{z}}{1-r_{z}},\\
r_{z} & \leq s_{z}\leq1.
\end{align}
In the left part of Fig.~\ref{fig:4} we show this region for the
initial state with the Bloch vector $\boldsymbol{r}=(0.5,0,0.5)^{T}$.

We will now consider another special case, namely $t_{z}=-1$. In
this situation we see from Eq.~(\ref{eq:constraint}) that $|b_{1}|=1$,
and $a_{2}^{2}$ can take any value between 0 and 1. From Eq.~(\ref{eq:Sz})
we get 
\begin{equation}
s_{z}=r_{z}-a_{2}^{2}(1+r_{z}),\label{eq:Sz-3}
\end{equation}
which implies that $s_{z}$ can take any value between $-1$ and $r_{z}$.
Moreover, for $|b_{1}|=1$ and a fixed value of $|a_{2}|$ the maximal
value of $s_{x}$ is given as $s_{x}=|r_{x}| ({1-a_{2}^{2}} )^{1/2}$,
which can be seen directly from Eq.~(\ref{eq:Sx}). Together with
Eq.~(\ref{eq:Sz-3}) we arrive at the following result for the maximal
value of $s_{x}$: 
\begin{equation}
s_{x}=\left|r_{x}\right|\sqrt{\frac{1+s_{z}}{1+r_{z}}}.
\end{equation}
By symmetry, this proves that the achievable region in the $x$-$z$
plane is determined by 
\begin{align}
\frac{s_{x}^{2}}{r_{x}^{2}} & \leq\frac{1+s_{z}}{1+r_{z}},\\
-1 & \leq s_{z}\leq r_{z}.
\end{align}
In general, i.e., without restricting to the $x$-$z$ plane, the
achievable region is given by
\begin{align}
\frac{s_{x}^{2}+s_{y}^{2}}{r_{x}^{2}+r_{y}^{2}} & \leq\frac{1+s_{z}}{1+r_{z}},\\
-1 & \leq s_{z}\leq r_{z}.
\end{align}
In the right part of Fig.~\ref{fig:4} we show this region for the
initial state with the Bloch vector $\boldsymbol{r}=(0.5,0,0.5)^{T}$.

In the final step we consider the remaining case $-1<t_{z}<1$, $t_{z}\neq0$.
By solving Eq.~(\ref{eq:constraint}) for $|b_{1}|^{2}$ and inserting
it in Eq.~(\ref{eq:Sz}) we obtain
\begin{align}
s_{z} & =r_{z}+2\left(\frac{t_{z}-r_{z}}{1-t_{z}}\right)a_{2}^{2}.\label{eq:Sz-1}
\end{align}
Using this result together with Eq.~(\ref{eq:a2}) we can determine
the range of $s_{z}$, given by
\begin{equation}
\begin{aligned}s_{z} & \in\left[r_{z},r_{z}+2\left(\frac{t_{z}-r_{z}}{1-t_{z}}\right)\times\min\left\{ 1,\frac{1-t_{z}}{1+t_{z}}\right\} \right]\,\,\,\mathrm{for}\,\,t_{z}\geq r_{z},\\
s_{z} & \in\left[r_{z}-2\left(\frac{r_{z}-t_{z}}{1-t_{z}}\right)\times\min\left\{ 1,\frac{1-t_{z}}{1+t_{z}}\right\} ,r_{z}\right]\,\,\,\mathrm{for}\,\,t_{z}<r_{z}.
\end{aligned}
\end{equation}
We will now determine all possible values of $s_{x}$ for a given
value of $s_{z}$. Note that due to Eqs.~(\ref{eq:constraint}) and
(\ref{eq:Sz-1}) a fixed value of $s_{z}$ immediately fixes the values
of $|a_{2}|$ and $|b_{1}|$ as follows: \begin{subequations} \label{eq:a2b1}
\begin{align}
|a_{2}| & =\sqrt{\frac{(s_{z}-r_{z})(1-t_{z})}{2(t_{z}-r_{z})}},\\
|b_{1}| & =\sqrt{1-\frac{(1+t_{z})(s_{z}-r_{z})}{2(t_{z}-r_{z})}}.
\end{align}
\end{subequations} In the next step, we use the explicit expression
for $s_{x}$ in Eq.~(\ref{eq:Sx}). Recalling that the vectors $\boldsymbol{a}=(a_{1},a_{2},a_{3})$
and $\boldsymbol{b}=(b_{1},b_{2},b_{3})$ are normalized, it is straightforward
to see that for fixed values of $|a_{2}|$ and $|b_{1}|$ the maximal
value for $s_{x}$ is given by
\begin{equation}
s_{x}=\left|r_{x}\right|\left(\left|b_{1}\right|\sqrt{1-a_{2}^{2}}+\left|a_{2}\right|\sqrt{1-\left|b_{1}\right|^{2}}\right).\label{eq:Sx-1}
\end{equation}
By using Eqs.~(\ref{eq:a2b1}) it is straightforward to express $s_{x}$
as a function of $r_{x}$, $r_{z}$, $s_{z}$, and $t_{z}$. We do
not display the final expression here. Recalling that the achievable
region is symmetric under rotation around the $z$-axis, it is straightforward
to obtain the full achievable region from Eq.~(\ref{eq:Sx-1}). 
In the left part of Fig.~\ref{fig:3} we show the achievable region
for the initial state with the Bloch vector $\boldsymbol{r}=(0.5,0,0.5)^{T}$,
and the state $\tau$ has the Bloch vector $\boldsymbol{t}=(0,0,0.7)^{T}$.

\section{\label{sec:Proof-Finite}Proof of Proposition \ref{prop:finite}}

We will present the proof for IO. The proof for SIO follows the same
lines of reasoning. In the first step we define the operator 
\begin{equation}
M=(K\otimes\openone)\Phi_{d}^{+}(K^{\dagger}\otimes\openone),\label{eq:M}
\end{equation}
where $K$ is an arbitrary incoherent operator. The operators $M$
are Hermitian operators of dimension $d^{2}$, i.e., the number of
real parameters is $d^{4}$. For any incoherent operation $\Lambda$,
the corresponding Choi state $\rho_{\Lambda}$ belongs to the convex
hull of the operators $M$. This follows directly from the definition
of an incoherent operation in Eq.~(\ref{eq:Lambda}). By Caratheodory's
theorem, any Choi state can thus be written as a convex combination
of at most $n=d^{4}+1$ operators of the form~(\ref{eq:M}). That
is, there exist $n$ incoherent operators $N_{j}$ and a probability
distribution $p_{j}$ such that 
\begin{equation}
\rho_{\Lambda}=\sum_{j=1}^{n}p_{j}\left(N_{j}\otimes\openone\right)\Phi_{d}^{+}(N_{j}^{\dagger}\otimes\openone).\label{eq:Choi state}
\end{equation}
This, together with the Choi-Jamio\l kowski isomorphism implies that
the incoherent operation $\Lambda$ can be written as 
\begin{equation}
\Lambda(\sigma)=\sum_{j=1}^{n}L_{j}\sigma L_{j}^{\dagger}
\end{equation}
with $n$ incoherent Kraus operators defined as $L_{j}=\sqrt{p_{j}}N_{j}$.
The fact that $L_{j}$ are indeed Kraus operators, i.e., fulfill the
completeness relation $\sum_{j=1}^{n}L_{j}^{\dagger}L_{j}=\openone$
can be checked directly from Eq.~(\ref{eq:Choi state}), recalling
that $\mathrm{Tr}_{1}(\rho_{\Lambda})=\openone/d$.

\section{\label{sec:Proof-Qubit}Proof of Proposition \ref{prop:qubit}}

The main ingredient in the proof is the fact that two sets of Kraus
operators $\{K_{j}\}_{j=1}^{n}$ and $\{L_{i}\}_{i=1}^{k}$ give rise
to the same quantum operation if and only if \cite{Nielsen2010} 
\begin{equation}
L_{i}=\sum_{j=1}^{n}U_{i,j}K_{j}\label{Eq:equivalence.of.Kraus.operators}
\end{equation}
where $U_{i,j}$ are the elements of a unitary $U\in U(\max\{n,k\})$.
That is to say, the two sets of Kraus operators are connected by an
isometry. The smallest number of Kraus operators achievable for a
given quantum operation is the \emph{Kraus rank}, which is identical
with the rank of the Choi state.

Before we proceed, we note that any incoherent operation on a finite
dimensional Hilbert space admits a finite set of incoherent Kraus
operators, as was proven in Proposition~\ref{prop:finite}. For a
single qubit, the incoherence condition $K_{j}\ket{m}\sim\ket{n}$
implies that every incoherent Kraus operator belongs to one of the
following four types: 
\begin{align}
K^{I} & =\left\{ \begin{pmatrix}* & *\\
0 & 0
\end{pmatrix}\right\} , & K^{II} & =\left\{ \begin{pmatrix}* & 0\\
0 & *
\end{pmatrix}\right\} ,\label{eq:types}\\
K^{III} & =\left\{ \begin{pmatrix}0 & 0\\
* & *
\end{pmatrix}\right\} , & K^{IV} & =\left\{ \begin{pmatrix}0 & *\\
* & 0
\end{pmatrix}\right\} ,
\end{align}
where {*} denotes an arbitrary complex number.

We will now show that there always exists a set of incoherent Kraus
operators $\{L_{i}\}$ where each of the four types occurs at most
twice, i.e., the total number of incoherent Kraus operators $L_{i}$
is at most eight. For this, assume that the incoherent Kraus representation
$\{K_{j}\}_{j=1}^{n}$ contains at least three nonzero Kraus operators
of the first type, i.e., $K_{1},K_{2},K_{3}\in K^{I}$. Note that
these three Kraus operators are then linearly dependent: there is
a nontrivial choice of numbers $z_{i}\in\mathbbm{C}$ such that $\sum_{i=1}^{3}z_{i}K_{i}=0$.
Without loss of any generality, we can assume that the vector $(z_{1},z_{2},z_{3})^{T}$
is normalized, i.e., $\sum_{i=1}^{3}|z_{i}|^{2}=1$. In the next step
we introduce a $U\in U(n)$ as 
\begin{equation}
U=V\oplus\openone_{n-3},
\end{equation}
where $\openone_{n-3}$ is the identity operator acting on dimension
$n-3$ and $V\in U(3)$ unitary defined such that its first row corresponds
to $(z_{1},z_{2},z_{3})$. With this definition, it is straightforward
to check that the Kraus operators $L_{i}=\sum_{j}U_{i,j}K_{j}$ are
all incoherent, and moreover $L_{1}=0$. These arguments are not limited
to Kraus operators of the first type $K^{I}$, but can be applied
in the same way for all types given above. Applying this procedure
repeatedly, we see that any incoherent single-qubit quantum operation
can be written with at most two incoherent Kraus operators of every
type, i.e, with at most $8$ incoherent Kraus operators in total.

In the next step, we show that the number of incoherent Kraus operators
can be further reduced to 6. For this, note that without any loss
of generality, the respective sets are given by 
\begin{align}
K^{I} & =\left\{ \begin{pmatrix}* & 0\\
0 & 0
\end{pmatrix},\,\begin{pmatrix}* & *\\
0 & 0
\end{pmatrix}\right\} , & K^{II} & =\left\{ \begin{pmatrix}* & 0\\
0 & 0
\end{pmatrix},\,\begin{pmatrix}* & 0\\
0 & *
\end{pmatrix}\right\} ,\\
K^{III} & =\left\{ \begin{pmatrix}0 & 0\\
* & 0
\end{pmatrix},\,\begin{pmatrix}0 & 0\\
* & *
\end{pmatrix}\right\} , & K^{IV} & =\left\{ \begin{pmatrix}0 & 0\\
* & 0
\end{pmatrix},\,\begin{pmatrix}0 & *\\
* & 0
\end{pmatrix}\right\} .
\end{align}
Following similar arguments as above, the joint set $K^{I}\cup K^{II}$
can be reduced to three operators, and the same is true for $K^{III}\cup K^{IV}$.
This proves that any incoherent single-qubit operation can be written
with 6 Kraus operators of the form 
\begin{equation}
\left\{ \begin{pmatrix}* & *\\
0 & 0
\end{pmatrix},\begin{pmatrix}0 & 0\\
* & *
\end{pmatrix},\begin{pmatrix}* & 0\\
0 & *
\end{pmatrix},\begin{pmatrix}0 & *\\
* & 0
\end{pmatrix},\begin{pmatrix}* & 0\\
0 & 0
\end{pmatrix},\begin{pmatrix}0 & 0\\
* & 0
\end{pmatrix}\right\} .
\end{equation}

We will now complete the proof, showing that this set of 6 Kraus operators
can be reduced to 5 Kraus operators of the form~(\ref{eq:IO-qubit}).
For achieving this, we will rearrange the aforementioned Kraus operators,
focusing in particular on the 3 operators 
\begin{equation}
K_{1}=\left(\begin{array}{cc}
0 & 0\\
a_{1} & b_{1}
\end{array}\right),\,\,\,K_{2}=\left(\begin{array}{cc}
a_{2} & 0\\
0 & b_{2}
\end{array}\right),\,\,\,K_{3}=\left(\begin{array}{cc}
0 & 0\\
a_{3} & 0
\end{array}\right),
\end{equation}
where $a_{i}$ and $b_{i}$ are complex numbers. The missing 3 Kraus
operators will be denoted by $K_{4}$, $K_{5}$, and $K_{6}$. Their
elements will not be important in the following discussion, but we
note that they have the form 
\begin{equation}
K_{4}=\left(\begin{array}{cc}
* & *\\
0 & 0
\end{array}\right),\,\,\,K_{5}=\begin{pmatrix}0 & *\\
* & 0
\end{pmatrix},\,\,\,K_{6}=\begin{pmatrix}* & 0\\
0 & 0
\end{pmatrix},\label{eq:Remaining}
\end{equation}
where $*$ denotes some complex number.

Consider now the following $3\times3$ unitary matrix 
\begin{equation}
U=\left(\begin{array}{ccc}
la_{1}^{*} & 0 & la_{3}^{*}\\
mb_{1}^{*}|a_{3}|^{2} & m\left(\left|a_{1}\right|{}^{2}+\left|a_{3}\right|{}^{2}\right)b_{2}^{*} & -ma_{3}^{*}b_{1}^{*}a_{1}\\
na_{3}b_{2} & -na_{3}b_{1} & -na_{1}b_{2}
\end{array}\right),\label{eq:U}
\end{equation}
where the parameters $l$, $m$, and $n$ are nonnegative and chosen
as 
\begin{eqnarray*}
l^{2} & = & \frac{1}{\left|a_{1}\right|{}^{2}+\left|a_{3}\right|{}^{2}},\\
m^{2} & = & \frac{1}{\left(\left|a_{1}\right|{}^{2}+\left|a_{3}\right|{}^{2}\right)\left(\left|a_{3}\right|{}^{2}\left[\left|b_{1}\right|{}^{2}+\left|b_{2}\right|{}^{2}\right]+\left|a_{1}b_{2}\right|{}^{2}\right)},\\
n^{2} & = & \frac{1}{\left|a_{3}\right|{}^{2}\left(\left|b_{1}\right|{}^{2}+\left|b_{2}\right|{}^{2}\right)+\left|a_{1}b_{2}\right|{}^{2}}.
\end{eqnarray*}
We now introduce a new Kraus decomposition $\{L_{i}\}$ as 
\begin{equation}
L_{i}=\begin{cases}
\sum_{j=1}^{3}U_{i,j}K_{j} & \mathrm{for}\,1\leq i\leq3,\\
K_{i} & \mathrm{for}\,4\leq i\leq6,
\end{cases}
\end{equation}
where $U_{i,j}$ are elements of the unitary matrix in Eq.~(\ref{eq:U}).
It can now be verified by inspection that the operators $L_{1}$,
$L_{2}$, and $L_{3}$ have the form 
\begin{equation}
L_{1}=\left(\begin{array}{cc}
0 & 0\\
* & *
\end{array}\right),\,\,\,L_{2}=\begin{pmatrix}* & 0\\
0 & *
\end{pmatrix},\,\,\,L_{3}=\begin{pmatrix}* & 0\\
0 & 0
\end{pmatrix},
\end{equation}
where $*$ denote some complex numbers which can be written in terms
of the parameters $a_{i}$ and $b_{i}$, but the explicit form is
not important for the following discussion.

Recall now that the remaining Kraus operators have the same form as
in Eq.~(\ref{eq:Remaining}), i.e., 
\begin{equation}
L_{4}=\left(\begin{array}{cc}
* & *\\
0 & 0
\end{array}\right),\,\,\,L_{5}=\begin{pmatrix}0 & *\\
* & 0
\end{pmatrix},\,\,\,L_{6}=\begin{pmatrix}* & 0\\
0 & 0
\end{pmatrix},
\end{equation}
and in particular the operator $L_{6}$ has the same form as $L_{3}$.
Thus, the set $L_{3}\cup L_{6}$ can be reduced to one Kraus operator.
This proves that any incoherent operation on a single qubit can be
written with at most 5 Kraus operators as given in Eq.~(\ref{eq:IO-qubit}). 

From the completeness condition $\sum_{i=1}^{5}K_{i}^{\dagger}K_{i}=\openone$,
it is straightforward to see that $\sum_{i=1}^{5}|a_{i}|^{2}=\sum_{j=1}^{4}|b_{j}|^{2}=1$
and $a_{1}^{*}b_{1}+a_{2}^{*}b_{2}=0$. Finally, the numbers $a_{j}=|a_{j}|e^{i\phi_{j}}$
can be chosen real by multiplying each of the Kraus operators with
the corresponding phase $e^{-i\phi_{j}}$.

\section{\label{sec:Proof-General}Proof of Proposition~\ref{prop:general}}

The proof follows similar ideas as the proof of Proposition~\ref{prop:qubit}.
We start with the fact that every incoherent operation admits a finite
decomposition into incoherent Kraus operators, see Proposition~\ref{prop:finite}.
Now we classify the incoherent Kraus operators according to their
types, i.e., the location of their nonzero elements. In general, there
are $d^{d}$ different types. For each type, at most $d$ incoherent
Kraus operators can be linearly independent. Following the arguments
in the proof of Proposition~\ref{prop:qubit}, this implies that
any incoherent operation admits a decomposition with at most $d^{d+1}$
incoherent Kraus operators.

To reduce this number further, let $C_{k}$ denote the number of incoherent
Kraus operators such that the first $k-1$ columns have all entries
zero, and the $k$-th column has one nonzero entry. In particular,
$C_{1}$ denotes the number of Kraus operators which have one nonzero
element in the first column, and possibly some nonzero elements in
the other columns. Moreover, $C_{d}$ denotes the number of Kraus
operators which have one nonzero entry in the last column, and all
other elements are zero.

As we will show in the following, every incoherent operation admits
an incoherent Kraus decomposition such that 
\begin{equation}
C_{k}\leq d^{d-k+1}\label{eq:Ck}
\end{equation}
for all $k\in[1,d]$. For proving the statement, note that $d^{d-k+1}$
is exactly the number of different shapes for incoherent Kraus operators
which have zero entries in the first $k-1$ columns. Assume now that
some Kraus decomposition $\{K_{i}\}$ has $C_{k}>d^{d-k+1}$ . Then,
there must be two Kraus operators $K_{1}$ and $K_{2}$ which have
the same shape and all entries in the first $k-1$ columns are zero.
Then -- similar as in the proof of Proposition~\ref{prop:qubit}
-- we can introduce a new incoherent Kraus decomposition $\{L_{i}\}$
such that $L_{1}$ and $L_{2}$ are linear combinations of $K_{1}$
and $K_{2}$. Moreover, we can choose the new Kraus operators such
that $L_{i}=K_{i}$ for $i>2$, and all elements in the $k$-th column
of $L_{1}$ are zero. This proves the existence of an incoherent Kraus
decomposition fulfilling Eq.~(\ref{eq:Ck}).

To complete the proof of the proposition we evaluate the sum over
all $C_{k}$, giving rise to 
\begin{equation}
\sum_{k=1}^{d}C_{k}\leq\sum_{n=1}^{d}d^{n}=\frac{d(d^{d}-1)}{d-1}.
\end{equation}

\section{\label{sec:Proof-prop:SIO}Proof of Proposition \ref{prop:SIO}}

A strictly incoherent Kraus operator can have at most one nonzero
element in each row and column. Similar as in the proof of Proposition~\ref{prop:general},
we now introduce the number $C_{k}$, which counts the number of strictly
incoherent Kraus operators with the first $k-1$ columns having all
entries zero, and the $k$-th column having one nonzero entry. We
will now show that every SIO admits a strictly incoherent Kraus decomposition
such that 
\begin{equation}
C_{k}\leq\frac{d!}{(k-1)!}.\label{eq:Ck-SI}
\end{equation}
For this, note that ${d!}/{(k-1)!}$ is the number of different shapes
of strictly incoherent Kraus operators which have zero entries in
the first $k-1$ columns. Assume now that some strictly incoherent
Kraus decomposition $\{K_{i}\}$ has $C_{k}>{d!}/{(k-1)!}$. Then,
there must be two Kraus operators $K_{1}$ and $K_{2}$ which have
the same shape and all entries in the first $k-1$ columns are zero.
We can then introduce a new Kraus decomposition $\{L_{i}\}$ such
that $L_{1}$ and $L_{2}$ are linear combinations of $K_{1}$ and
$K_{2}$. Moreover, we can choose the new Kraus operators such that
$L_{i}=K_{i}$ for $i>2$, and all elements in the $k$-th column
of $L_{1}$ are zero. This proves the existence of a strictly incoherent
Kraus decomposition which fulfills Eq.~(\ref{eq:Ck-SI}). The proof
of the proposition is complete by noticing that the sum $\sum_{k=1}^{d}C_{k}$
is an upper bound on the total number of strictly incoherent Kraus
operators, and thus 
\begin{equation}
\sum_{k=1}^{d}C_{k}\leq\sum_{k=1}^{d}\frac{d!}{(k-1)!}.
\end{equation}

\section{\label{sec:Lower-Bound}Proof of Proposition \ref{prop:SIO2}}

Consider a quantum operation defined via the Kraus operators 
\begin{equation}
K_{i,j}=\frac{1}{\sqrt{d}}\ket{\mathrm{mod}(j+i,d)}\!\bra{j}
\end{equation}
with $0\leq i,j\leq d-1$. We will now prove that the strictly incoherent
operation $\Lambda(\rho)=\sum_{i,j}K_{i,j}\rho K_{i,j}$ cannot be
implemented with fewer than $d^{2}$ Kraus operators. For this, it
is enough to show that the corresponding Choi state $\rho_{\Lambda}$
has rank $d^{2}$. This can be seen by applying the operators $K_{i,j}\otimes\openone$
to the maximally entangled state vector $\ket{\Phi_{d}^{+}}=\sum_{k=0}^{d-1}\ket{k,k}/\sqrt{d}$:
\begin{eqnarray}
\left(K_{i,j}\otimes\openone\right)\ket{\Phi_{d}^{+}} & = & \frac{1}{d}\ket{\mathrm{mod}(j+i,d)}\ket{j}.
\end{eqnarray}
The proof is complete by observing that all these $d^{2}$ (unnormalized)
states are linearly independent. 
\end{document}